\documentstyle[12pt]{article}
\def\beq{\begin{equation}}   \def\eeq{\end{equation}}
\def\bea{\begin{eqnarray}}   \def\eea{\end{eqnarray}}

\begin{document}

\begin{flushright}
UND-HEP-00-BIG\hspace*{.2em}10\\
%hep-ph/0011231\\
%{\footnotesize leapv2.tex}~~~~\today  \\ 
%version  2.0\\ 
\end{flushright}
\vspace{.3cm}
\begin{center} \Large 
{\bf CP, T and CPT Symmetries at the Turn of a New 
Millenium}
\footnote{Lecture given at LEAP2000, Venice, 
Italy, Aug. 21 - 26, 2000}
\\
\end{center}
\vspace*{.3cm}
\begin{center} {\Large 
I. I. Bigi }\\ 
\vspace{.4cm}
{\normalsize 
{\it Physics Dept.,
Univ. of Notre Dame du
Lac, Notre Dame, IN 46556, U.S.A.} }
\\
\vspace{.3cm}
{\it e-mail address: bigi@undhep.hep.nd.edu } 
\vspace*{0.4cm}

{\Large{\bf Abstract}}
\end{center}

After summarizing the status concerning CP violation in 1998 I 
describe the exciting developments of the last two 
years and extrapolate to the future. I comment on recent lessons 
about T and CPT invariance maninly from CPLEAR and emphasize the 
potential of finding New Physics by analyzing  $K_{\mu 3}$ 
and charm decays and searching for electric dipole moments.

%\vspace*{.4cm}
%\vfill
%\noindent
%\vskip 5mm
%PACS 11.30.Er, 13.20.Eb, 13.25.Es
%\vskip 3mm

%]
%%%%%%%%%%%%
\tableofcontents 
%%%%%%%%%%

\vspace*{1.0cm}

%%%%%%%%%%%%%%%
\section{Introduction}
%%%%%%%%%%%%%

During my talk I want to focus on three topics, namely 
\begin{itemize}
\item 
the exciting developments in heavy flavour physics of the last two 
years and what can be expected in the foreseeable future, 
\item 
the lessons on T and CPT invariance learnt from CPLEAR as well as 
KTeV and NA48 including some comments on a program for the AD program 
at CERN, and 
\item 
other non-mainstream trends.

\end{itemize}

More specifically my talk will be organized as follows: after 
reminding you of why CP violation represents such a 
fundamental phenomenon and sketching the CP phenomenology 
as it existed in 1998 in Sect.2, I will describe the new 
insights and developments since then and what can be expected 
in the next decade or so in Sect.3; in Sect.4 I discuss T and 
CPT invariance and what has been learnt about it from CPLEAR with 
some additional information from $K_L \to \pi ^+\pi ^- e^+e^-$; 
in Sect.5 I comment on `exotica', namely direct CP violation in 
hyperon decays, on $K_{\mu 3}$ decays, electric dipole moments 
and CP violation in charm transitions before presenting an outlook 
in Sect.6. 

%%%%%%%%%%%%%%%%%%%%%%%%
\section{CP invariance and its limitations through 1998}
%%%%%%%%%%%%%%%%  

%%%%%%%%%%%%
\subsection{CP violation as a fundamentally new paradigm}
%%%%%%%%%%

The discovery in 1957 that parity was broken in weak decays 
certainly caused a shock in the community. Yet the latter   
recovered remarkably fast largely due to arguments put 
forward by leading physicists like Landau. They suggested 
one had been hasty in requiring full invariance under parity.  
Invoking somewhat obliquely Mach's principle they 
instead argued in 
favour of CP symmetry 
pairing {\em left}-handed neutrinos 
with {\em right}-handed {\em anti}neutrinos; `left' and `right' 
is then defined in terms `positive' and `negative'. 
\footnote{This is 
similar to a German saying that the thumb is `left' on the 
`right' hand: it is as factually correct as it is useless 
since circular.} 
A world of left-handed fermions and right-handed 
antifermions is thus a completely symmetric one. Indeed it was 
found that maximal parity violation in weak 
interactions is balanced by maximal 
violation of charge conjugation. This might remind you of the 
literary figure of `a man without a future and a woman 
without a past'. 

The observation of $K_L \to \pi ^+\pi ^-$ in 1964 was totally 
unexpected by almost all theorists, and they did not give up 
without a fight. Interpretations other than CP violation 
were entertained: the existance of a particle $U$ 
escaping detection in $K_L \to \pi ^+\pi ^- [U]$ was postulated
\footnote{It is an argument analogous to Pauli's introduction 
of neutrinos into $\beta$ decay: 
an `invisible' particle is postulated 
to save a conservation law, namely that of energy-momentum 
there and CP here. While this idea worked there, it failed here.} 
; 
cosmological background fields were invoked and even the idea 
of {\em nonlinear effects} in quantum mechanics were floated 
\cite{ROOS}  
-- to no avail! 
The fact that CP invariance appeared to be a `near-miss' -- 
BR$(K_L \to \pi ^+\pi ^-) \sim 0.002 \ll 1$   
in contrast to maximal P violation --   
made it even harder to accept. 
Nevertheless the whole community soon came 
around to accept CP violation as an empirical fact 
\cite{PAIS,CPBOOK}. 

I am telling this story not to poke fun at my predecessors. 
There were very good reasons for theorists' slowness in embracing 
CP violation. For it was clearly realized that CP violation represented 
a more fundamental and radical shift to a new paradigm than 
parity violation. 
\begin{itemize}
\item 
CP violation means that 
`left'- and `right'-handed can be distinguished in an 
{\em absolute}  
way, independant of any convention concerning the sign of charges. 
This is most obvious from the observation on semileptonic 
$K_L$ decays:
\beq 
\Gamma (K_L \to l^+ \nu _L\pi^-) > 
\Gamma (K_L \to l^- \bar \nu _R\pi^+) \; . 
\eeq  
\item 
Due to CPT symmetry CP violation implies T violation, 
i.e. that nature distinguishes between `past' and `future' on the 
{\em microscopic} level. 
\item 
One can add (at least in retrospect) 
 that CP violation is a necessary ingredient 
in any effort to understand the baryon number of the Universe 
as a {\em dynamically generated} quantity rather than as a parameter 
reflecting {\em initial conditions}. 
\item 
On a more technical level one 
can point out that CP violation represents the smallest observed 
violation of a symmetry:  
Im$M_{12} \simeq 1.1 \cdot 10^{-8}$ eV or 
Im$M_{12}/m_K \simeq 2.2 \cdot 10^{-17}$. 
\item 
The peculiar role 
of T violation surfaces also through {\em Kramers' Degeneracy} 
\cite{KRAMERS}. 
With the time reversal operator $T$ being {\em anti}unitary, 
$T^2$ has eigenvalues $\pm 1$ meaning the Hilbert space has two 
distinct sectors. It is easily shown that each energy eigenstate 
in the sector with $T^2 = -1$ is at least doubly degenerate. 
This degeneracy is realized in nature through fermionic degrees 
of freedom. I find it quite remarkable that the operator $T$ 
anticipates this option (and the qualitative difference between 
fermions and bosons) through $T^2=-1$ {\em without} any explicit 
reference to spin.  

\end{itemize}

%%%%%%%%%%%
\subsection{Basic CP [\& T] phenomenology}
%%%%%%%%%%%%

Due to CPT symmetry CP and T violation can enter through complex phases 
only. For them to become observable, one needs two different 
amplitudes to contribute coherently. This can be realized 
in different ways: 
\begin{itemize}
\item 
{\em Particle-antiparticle oscillations followed by a decay into 
a common final state:} 

Such asymmetries are often referred to -- with less than Shakespearean 
flourish -- as indirect CP violation. 
The decay rate evolution in proper time then 
differs from a {\em pure exponential}, and the 
difference between CP conjugate transitions becomes a nontrivial 
function of time. Well-known examples are 
$K^0(t) \to \pi ^+\pi^-$ vs. $\bar K^0(t) \to \pi ^+\pi^-$ 
or $B_d(t) \to \psi K_S$ vs. $\bar B_d(t) \to \psi K_S$ with 
\cite{BS} 
\beq  
\Gamma (B_d(t)[\bar B_d(t)] \to \psi K_S) \propto 
e^{-t/\tau (B_d)} \left( 
1 - [+] A {\rm sin}(\Delta m_d t)\right) 
\eeq
Final state interactions (FSI) in general will affect the 
signal, although not for $B_d \to \psi K_S$. 
On the other hand they are not required and they cannot fake a signal. 
\item 
{\em Direct CP  violation:} 

Within the SM 
it can occur in CKM suppressed modes only.
There are several classes of such effects 
differing in the role played by final state interactions; 
they all share the feature 
that the signal is independant of the time of decay. 
\begin{itemize}
\item 
{\em Partial width differences:} 
The prime example is provided by comparing the strength of the 
two CP violating transitions $K_L \to \pi ^+ \pi ^-$ and 
$K_L \to \pi ^0 \pi ^0$: 
\bea 
\eta _{+-} &\equiv& 
\frac{T(K_L\to \pi ^+ \pi ^-)}{T(K_S\to \pi ^+ \pi ^-)}
\equiv \epsilon + \epsilon ^{\prime} \\
\eta _{00} &\equiv& 
\frac{T(K_L\to \pi ^0 \pi ^0)}{T(K_S\to \pi ^0 \pi ^0)}
\equiv \epsilon - 2 \epsilon ^{\prime}
\eea 
While the quantity $\epsilon$ characterizes the decaying state 
$K_L$, $\epsilon ^{\prime}$ differentiates between the CP 
properties of the final states $\pi ^+\pi ^-$ versus 
$\pi ^0\pi^0$. Its value can be determined from decay rates: 
\beq 
{\rm Re} \frac{\epsilon ^{\prime}}{\epsilon } = 
\frac{1}{6} \left[ 
\frac{\Gamma (K_L\to \pi ^+\pi^-)/\Gamma (K_S\to \pi ^+\pi^-)}
{\Gamma (K_L\to \pi ^0\pi^0)/\Gamma (K_S\to \pi ^0\pi^0)} -1 \right] 
\eeq
This situation can be generalized. If the final 
state consists of two pseudoscalar mesons or one 
pseudoscalar and one vector meson, then CP violation can manifest 
itself only in a partial width difference.   
FSI are necessary to transform 
CP violation into an observable. While they cloud the 
numerical interpretation of a signal (or its absence), they cannot 
fake a signal.  
\item 
{\em Final state distributions:} If a final state is more complex, i.e. 
consists of at least three pseudoscalar mesons not forming  
a resonance or of two vector mesons etc., then there are several 
potential layers of dynamical information. There could be 
asymmetries in subregions of a Dalitz plot that are substantially larger 
than when integrated over the whole Dalitz plot. 

Going one step further one can study decays of a particle P into 
four pseudoscalar mesons: $P \to a+b+c+d$. Such a final state allows to 
construct non-trivial {\em T-odd} correlations:
\beq 
C_T\equiv 
\langle \vec p_a\cdot (\vec p_b \times \vec p_c)\rangle  
\eeq 
with $C_T \to - C_T$ under time reversal. T violation can 
produce $C_T \neq 0$ irrespective of FSI; yet 
$C_T \neq 0$ does not necessarily establish T violation. 
Since T is described by an {\em anti}unitary operator, FSI can 
induce $C_T \neq 0$ with T-invariant dynamics. 
In contrast to the situation with partial widths where 
FSI play the role of a necessary evil, here they can act as 
an imposter.  Yet comparing 
this observable for particle and antiparticle decays and finding 
$C_T + \bar C_T \neq 0$ establishes CP violation.  

The muon polarization transverse to the decay plane in 
$K^+ \to \mu ^+ \pi ^0 \nu$ represents such a T-odd correlation:  
$P_{\perp}(\mu ) = \langle \vec s(\mu ) \cdot 
(\vec p (\mu) \times \vec p (\pi))/
|\vec p (\mu) \times \vec p (\pi)| \rangle$, 
which 
in this case could not be faked realistically 
by final-state interactions and
would  reveal genuine T violation.

\item 
The leading, namely linear term for the energy shift of a system inside a  
weak electric field $\vec E$ is described by a static 
quantity, the electric dipole moment $\vec d$: 
\beq 
\Delta {\cal E} = \vec d \cdot \vec E + {\cal O}(E^2) 
\eeq
For a non-degenerate system with spin $\vec s$ one has 
$\vec d \propto \vec s$; therefore $\vec d \neq 0$ reveals 
T (and P) violation.

\end{itemize}

\end{itemize}

%%%%%%%%%%%%%%%
\subsection{Theory of CP violation}
%%%%%%%%%%%%%%

Initially it had been suggested that electrodynamics might 
violate CP invariance; yet it was soon cleared of that 
suspicion. There was then no theory of CP violation 
till 1972. The community can be forgiven for not being overly 
concerned about explaining BR$(K_L \to \pi ^+\pi ^-) 
\simeq 0.002$ when there are still infinities arising in the 
theoretical description of weak decays. Yet I find it highly 
remarkable that even after the SM had been formulated as a 
{\em renormalizable} theory by the late 1960's the lack of a theory 
for CP violation was not noticed till 1972 
\cite{MOHA}. It is often 
said in response:"Well, we had the superweak model put forward 
by Wolfenstein already in 1964". However I view the superweak 
model 
\cite{SUWE} as a {\em classification} scheme for theories 
rather than a theory 
itself. Whenever one suggests a theory of CP violation, one has to 
analyze whether it provides a dynamical implementation of the 
superweak scenario or not, and to which accuracy it does so. 

In 1973 the celebrated paper by Kobayashi and Maskawa appeared 
in print 
\cite{KM}. It pointed out that the electroweak SM with two full 
families -- i.e. charm included -- conserves CP; secondly it 
demonstrated how different types of New Physics -- more families, 
more Higgs doublets, right-handed currents -- allow CP breaking 
\footnote{It had been noted first by Mohapatra that the SM with two 
families conserves CP. He suggested right-handed currents as 
the origin of CP violation \cite{MOHA}.} 
\footnote{One can point out that Kobayashi and Maskawa benefitted from 
some `insider' information: both were working in the Physics 
Department of Nagoya University at that time where, due to 
Sakata and his school, the notion of quarks as 
real rather than merely mathematical objects had been 
readily accepted, as had been the existence of charm due to the 
discovery of Niu \cite{NIU}.}.   
Only one of these variants, namely the one with (at least) three 
families is now referred to as KM ansatz. 

This KM ansatz removes the mystery from the 
apparent `near miss' of CP invariance in $K_L \to \pi \pi$: this 
transition requires the interplay between three families; yet 
the third family is almost decoupled from the first two -- 
not surprisingly (again at least in retrospect) 
considering its much heavier masses. 

A second milestone was reached in the 1970's when the 
relevance of the so-called 
Penguin operators was realized, first in the context of the 
$\Delta I=1/2$ rule 
\cite{ITEP}, then also for allowing for 
$\epsilon ^{\prime}/\epsilon \neq 0$ 
\cite{GILMAN}. Since then the 
treatment of Penguin operators and operator renormalization 
has reached a high level of sophistication 
\cite{BURAS}.

A third milestone is represented by the formulation of the 
`Strong CP Problem'; it still awaits its resolution 
\cite{PECCEI}! 

Another milestone was the realization 
in 1980 that the KM ansatz unequivocally 
predicts large CP asymmetries in some nonleptonic decay 
channels of neutral $B$ mesons like $B_d \to \psi K_S$ 
\cite{CARTER,BS}. 
It was stated explicitely that 
asymmetries could be 1- 20 \% and possible larger -- at a 
time when neither the `long' $B$ lifetime nor the large 
$B_d - \bar B_d$ oscillation rate nor the huge top mass were 
known; at that time a top mass exceeding 60 GeV would have been seen 
as a frivolous notion! 

With only three families the unitarity constraints of the CKM 
matrix are conveniently expressed through triangle relations in the 
complex plane. The one most relevant for $B$ physics is given by 
\beq 
V^*(tb)V(td) + V^*(cb)V(cd) + V^*(ub)V(ud) = 0 
\label{THETRI}
\eeq
Various CP asymmetries are described in terms of the angles of this 
triangle; an ecumenical message in PDG2000 
endorses two different notations, namely 
$$ 
\phi _1 \equiv \beta = \pi - 
{\rm arg}\left( \frac{V(tb)^*V(td)}{V(cb)^*V(cd)}
\right) ,  
\phi _2\equiv \alpha =  
{\rm arg}\left( \frac{V(tb)^*V(td)}{-V(ub)^*V(ud)} 
\right) \; , 
$$
\beq  
\phi _3 \equiv \gamma =  
{\rm arg}\left( \frac{V(ub)^*V(ud)}{-V(cb)^*V(cd)}\right) . 
\eeq

%%%%%%%%%%%%%%%%%%%%
\subsection{The `unreasonable' success of the CKM description}
%%%%%%%%%%%%%%%%%%%%

The observation of the `long' $B$ lifetime of about 1 psec together 
with the dominance of $b\to c$ over $b\to u$ revealed a hierarchical 
structure in the KM matrix that is expressed in the Wolfenstein 
representation in powers of $\lambda = {\rm tg}\theta _C$. 
The triangle defined by Eq.(\ref{THETRI}) then takes on a very 
special form: its three sides are all of order 
$\lambda ^3$ and its angles therefore of order unity -- as are 
the CP asymmetries they describe! Details are given in 
Sect.\ref{EXPECT}. 

We   
often see plots of the CKM unitarity triangle where the 
constraints coming from various observables appear as    
broad bands. While the latter is often bemoaned, it obscures 
a more fundamental point: the fact that these constraints can be 
represented in such plots at all is quite amazing! 
The quark box 
{\em without} GIM subtraction yields a value for 
$\Delta m_K$ exceeding the experimental 
number by more than a factor of thousand; it is the GIM mechanism 
that brings it down to within a factor of two or so of experiment. 
The GIM subtracted quark box for $\Delta M_B$ coincides 
with the data again within a factor of two. Yet if the 
beauty lifetime were of order $10^{-14}$ sec while 
$m_t \sim 180$ GeV it would exceed it by 
an order of magnitude; on the other hand it would undershoot by an order 
of magnitude if $m_t \sim 40$ GeV were used with 
$\tau (B) \sim 10^{-12}$ sec; i.e., the observed value can be 
accommodated because a tiny value of $|V(td)V(ts)|$ is offset 
by a large $m_t$. 

This amazing success is repeated with $\epsilon$. Over the last 
25 years it could always be accommodated (apart from 
some very short periods of grumbling mostly off the record) 
whether the {\em correct} set [$m_t = 180$ GeV with $|V(td)|\sim 
\lambda ^3$, 
$|V(ts)|\sim \lambda ^2$] or the {\em wrong} one 
[$m_t = 40$ GeV with $|V(td)|\sim \lambda ^2$, 
$|V(ts)|=\lambda$] were used. Yet both 
$m_t = 180$ GeV with $|V(td)|=\lambda ^2$, 
$|V(ts)|=\lambda$ as well as 
$m_t = 40$ GeV with $|V(td)|=\lambda ^3$, 
$|V(ts)|=\lambda ^2$ would have lead to a clear inconsistency! 

Thus the phenomenological success of the CKM description has to be 
seen as highly nontrivial or `unreasonable'. This cannot have 
come about by accident -- there must be a profound reason.

%%%%%%%%%%%%%%%%%%%
\subsection{New QCD technologies of the 1990's}
%%%%%%%%%%%%%%%

Since we have to study the decays of quarks bound inside hadrons, 
we have to deal with nonperturbative dynamics 
\footnote{Since top quarks decay before they can hadronize, 
their interactions can be treated perturbatively 
\cite{DOK}.} 
-- 
a problem that in general has not been brought under theoretical 
control. Yet we can employ various theoretical technologies 
that allow to treat nonperturbative effects in special situation: 
\begin{itemize}
\item 
For {\em strange} hadrons where $m_s \leq \Lambda _{QCD}$ 
one invokes chiral perturbation theory. 
\item 
For {\em beauty} hadrons with $m_b \gg \Lambda _{QCD}$ one can 
employ $1/m_b$ expansions in various incarnations; they should provide 
us with rather reliable results, whenever an operator product expansion 
can be applied \cite{HQT}. 
\item 
It is natural to extrapolate such expansions down to the charm 
scale; this has to be done with considerable caution, though:  
while the charm quark mass exceeds ordinary hadronic mass 
scales, it does not do so by a large amount.  
\item 
Lattice QCD on the other hand is most readily set up at ordinary 
hadronic scales; from those one extrapolates {\em down} towards the chiral 
limit (which represents a nontrivial challenge) and 
{\em up} to the charm scale and beyond.

\end{itemize}

\noindent Let me add a few more specific comments: 
 
Lattice QCD, which originally had been introduced to prove confinement 
and bring hadronic spectroscopy under computational control is now making 
major contributions to heavy flavour physics. 
This can be illustrated with very recent results on decay constants 
where the first {\em un}quenched results (with two dynamical 
quark flavours) have become available \cite{KENWAY}. 
\begin{itemize}
\item 
\beq 
f(D_s) = 
\left\{ 
\begin{array}{l} 
240 \pm 4 \pm 24, \, 275 \pm 20 \;
{\rm MeV}, \; {\rm lattice\, QCD}\\ 
269 \pm 22 \; {\rm MeV}, \; {\rm world\, average\, of\, data}  
\end{array}  
\right. 
\eeq
\item 
\bea 
f(B) &=& 190 \pm 6 \pm 20 ^{+9}_{-0} \; {\rm MeV}, 
\; {\rm lattice\, QCD}\\
f(B_s) &=& 218 \pm 5 \pm 26 ^{+9}_{-0} \; {\rm MeV}, 
\; {\rm lattice\, QCD}
\eea
\end{itemize}

The $1/m_Q$ expansions have become more refined and reliable 
qualitatively as well as quantitatively: 
\begin{itemize}
\item 
The $b$ quark mass has been extracted 
from data by different groups; 
their findings, when expressed in terms of the 
socalled `kinetic' mass 
(which is distinct from both the pole as well as 
$\overline{{\rm MS}}$ 
mass), read as follows: 
\beq 
m_b^{\rm kin} (1\, {\rm GeV}) = 
\left\{ 
\begin{array}{l} 
4.56 \pm 0.06  \; \; 
{\rm GeV} \; \;  \cite{MEL}, \\  
4.57 \pm 0.04  \; \; 
{\rm GeV} \; \;  \cite{HOANG}, \\  
4.59 \pm 0.06  \; \; 
{\rm GeV} \; \;  \cite{SIGNER}
\end{array}  
\right.
\eeq
The error estimates of 1 - 1.5 \% might be overly optimistic (as it 
often happens), but not foolish. Since all 
three analyses use basically the same input from the 
$\Upsilon (4S)$ region, they could suffer from a common 
systematic uncertainty, though. 
\item 
For the form factor 
describing $B\to l \nu D^*$ at zero recoil
 one has the following results:
\beq 
F_{D^*}(0) = 
\left\{ 
\begin{array}{l} 
0.89 \pm 0.08  \; \;       \cite{URI1}, \\ 
0.913 \pm 0.042 \; \; \cite{BABARBOOK}, \\  
0.935 \pm 0.03 \; \; \cite{LAT} 
\end{array}  
\right. 
\eeq 
where the last number has been obtained in lattice QCD.  
\end{itemize}

There is a natural feedback between lattice QCD and $1/m_Q$ 
expansions: by now both represent mature technologies that 
are defined in Euclidean rather than Minkowskian space; 
they share some expansion parameters, while differing in others; 
lattice QCD can evaluate hadronic matrix elements that serve 
as input parameters to $1/m_Q$ expansions. 

It has been accepted for a long time now that heavy flavour decays 
can serve as high {\em sensitivity} probes for New Physics. I feel 
increasingly optimistic that our tools are and will be such that 
that they will provide us even with high {\em accuracy} probes!

%%%%%%%%%%%%%%%%%%%
\subsection{Expectations and predictions 1998 
\label{EXPECT}}
%%%%%%%%%%%%%%%%%

The observed hierarchy in the CKM parameters 
\beq 
|V(ub)|^2 \ll |V(cb)|^2 \ll |V(cd)|^2 
\eeq
tells us that the CKM matrix can conveniently be described by 
the Wolfenstein parametrization in powers of 
$\lambda = {\rm tg}(\theta _C)$: 
\beq 
V_{CKM} = \left( 
\begin{array}{ccc} 
V(ud) & V(us) & V(ub) \\ 
V(cd) &V(cs) & V(cb) \\ 
V(td) & V(ts) & V(tb) 
\end {array} 
\right) = 
\left( 
\begin{array}{ccc} 
1 & {\cal O}(\lambda ) & {\cal O}(\lambda ^3) \\ 
{\cal O}(\lambda ) & 1 & {\cal O}(\lambda ^2) \\ 
{\cal O}(\lambda ^3) & {\cal O}(\lambda ^2) & 1 
\end {array} 
\right) 
\label{WOLF} 
\eeq
More specifically PDG2000 states as 90\% C.L. ranges   
\beq 
|V_{CKM}| = 
\left( 
\begin{array}{ccc} 
0.9750 \pm 0.0008 & 0.223 \pm 0.004 & 0.003 \pm 0.002 \\ 
0.222 \pm 0.003 & 0.9742 \pm 0.0008 & 0.040 \pm 0.003 \\ 
0.009 \pm 0.005 & 0.039 \pm 0.004 & 0.9992 \pm 0.0002 
\end {array} 
\right)
\label{CKM3}
\eeq
Without imposing three-family unitarity that is implicit in the 
Wolfenstein representation PDG2000 lists numbers that in particular 
for the top couplings are much less restrictive:  
\beq 
|V_{CKM}| = 
\left( 
\begin{array}{cccc} 
0.9735 \pm 0.0013 & 0.220 \pm 0.004 & 0.003 \pm 0.002 & ...\\ 
0.226 \pm 0.007 & 0.880 \pm 0.096 & 0.040 \pm 0.003 & ...\\ 
0.05 \pm 0.04 & 0.28 \pm 0.27 & 0.5 \pm 0.49 & ... \\
... & ... & ... & ... 
\end {array} 
\right)
\label{CKM4}
\eeq
I would like to add three comments here: 
\begin{itemize} 
\item 
The brandnew CLEO number for $|V(cb)|$ from 
$B\to l \nu D^*$ -- 
$|V(cb)F_{D^*}(0)| = (42.4 \pm 1.8 \pm 1.9)\times 10^{-3}$ 
\cite{CINABRO} -- 
falls outside the 90\% C.L. range stated by PDG2000 
for the expected values of 
$F_{D^*}(0)$. 
\item 
The OPAL collaboration has presented 
a new {\em direct} determination of $|V(cs)|$ from 
$W\to H_c X$: $|V(cs)| = 0.969 \pm 0.058$ 
\cite{OPAL}. 
\item 
Using these values one finds 
\beq 
|V(ud)|^2 + |V(us)|^2 + |V(ub)|^2 = 1.000 \pm 0.003 \; , 
\eeq
which is perfectly consistent with the unitarity of the 
CKM matrix. Yet using instead $|V(ud)| = 
0.9740 \pm 0.0005$ as extracted from 
nuclear $0^+ \to 0^+$ transitions, one obtains \cite{TOWNER} 
\beq 
|V(ud)|^2 + |V(us)|^2 + |V(ub)|^2 = 0.9968 \pm 0.0014 \; , 
\eeq 
i.e., a bit more than a 2 $\sigma$ deficit in the unitarity 
condition. 
\end{itemize}

With these input values one can make predictions on CP asymmetries, 
at least in principle and to some degree. I will confine myself 
to a few more qualitative comments.  
\begin{itemize}
\item 
If there is a single CP violating phase 
$\delta$ as is the case in the KM 
ansatz one can conclude based on the $\Delta I = 1/2$ rule: 
$\epsilon ^{\prime}/\epsilon \leq 1/20$. The large top mass 
-- $m_t  \gg M_W$ -- enhances the SM prediction for $\epsilon$ 
considerably more than for $\epsilon ^{\prime}$ for a 
given $\delta$ and therefore on quite general grounds 
\beq 
\epsilon ^{\prime}/\epsilon \ll 1/20
\eeq
\item 
Of course the KM predictions made employed much more 
sophisticated theoretical reasoning. Before 1999 they tended to 
yield -- with few exceptions 
\cite{FABB} -- values not exceeding $10^{-3}$ 
due to sizeable cancellations between different contributions. 
\item 
Once the CKM matrix exhibits the {\em qualitative} 
pattern given in Eq.(\ref{WOLF}), 
it neccessarily follows that certain $B_d$ decay channels will 
exhibit CP asymmetries of order unity. To be more specific one can 
combine what is known about $V(cb)$, $V(ub)$, $V(ts)$ and $V(td)$ 
from semileptonic $B$ decays, $B_d - \bar B_d$ oscillations and 
bounds on $B_s - \bar B_s$ oscillations with or without using 
$\epsilon$ to construct the CKM unitarity triangle 
describing $B$ decays. A crucial question to which I will return 
later centers on the proper treatment of theoretical uncertainties. 
A typical example is \cite{PARODI}: 
\bea 
{\rm sin} 2\phi _1[\beta] &=& 0.716 \pm 0.070 \\ 
{\rm sin} 2\phi _2[\alpha] &=& - 0.26 \pm 0.28 
\eea 
\end{itemize}

%%%%%%%%%%%%%%%%%%%%%
\subsection{Status of the data in 1998}
%%%%%%%%%%%%%%%%%%%%

The relevant data read as follows in 1998: 
\begin{itemize}
\item 
\bea 
{\rm BR}(K_L \to \pi ^+ \pi ^-) &\simeq & 2.3 \cdot 10^{-3} \neq 0 \\
\frac{{\rm BR}(K_L \to l^+ \nu \pi ^-)}
{{\rm BR}(K_L \to l^- \nu \pi ^+)} &\simeq & 1.006 \neq 0
\eea
\item 
\beq 
{\rm Re} \frac{\epsilon ^{\prime}}{\epsilon _K} = 
\left\{ 
\begin{array}{l} 
(2.30 \pm 0.65) \cdot 10^{-3} \; \; NA\, 31 \\ 
(0.74 \pm 0.59) \cdot 10^{-3} \; \; 
E\, 731 
\end{array}  
\right. 
\label{DIRECTCP} 
\eeq
\item 
The muon transverse polarization in 
$K^+ \to \mu ^+ \nu \pi ^0$: 
\beq 
{\rm Pol}_{\perp} (\mu )   
 = (-1.85 \pm 3.6) \cdot 10^{-3} 
\eeq 
\item 
Electric dipole moments for neutrons and electrons 
\bea 
d_N &<& 9.7 \cdot 10^{-26} \; \; e\, cm \\
d_e &=& (-0.3 \pm 0.8) \cdot 10^{-26} \; \; e \, cm
\eea 
To get an intuitive understanding about the sensitivity 
achieved one can point out that the uncertainty in the 
electron's {\em magnetic} moment is about 
$2 \cdot 10^{-22}$ e cm and thus several orders of magnitude 
larger than the bound on its EDM! The bound on the neutron's 
EDM is smaller than its radius by 13 orders of magnitude. 
This corresponds to a relative displacement of an electron and a 
positron spread over the whole earth by less than 1 $\mu$ -- much 
less than the thickness of human hair!

\end{itemize}
The situation in 1998 can then be described as follows: after 34 
years of dedicated experimental work CP violation could still be 
described by a {\em single} number, namely $\epsilon$, the 
situation concerning direct CP violation was in limbo, see 
Eq.(\ref{DIRECTCP}), and no other manifestation had been seen.

%%%%%%%%%%%%%%%%%%%%
\section{New insights from 1999 and future developments}
%%%%%%%%%%%%%%%%%% 

Direct CP violation has been established 
in $K_L$ decays:
\beq 
{\rm Re}\left( \frac{\epsilon ^{\prime}}{\epsilon ^{\prime}}
\right) = 
\left\{ 
\begin{array}{l} 
(2.80 \pm 0.41)\cdot 10^{-3} \; \; \; {\rm KTeV}, \\   
(1.40 \pm 0.43)\cdot 10^{-3} \; \; \; {\rm NA48} \; ; 
\end{array}  
\right.
\eeq 
its exact size, however, is still uncertain. 
It is a discovery of the first rank irrespective of what 
theory says or does not say.

Our theoretical interpretation of the 
data is very much in limbo. As I had argued before a rather small, but 
nonzero value is a natural expectation of the KM ansatz. 
To go beyond such a qualitative statement, one has to evaluate 
hadronic matrix elements; apparently one had 
underestimated the complexities in this task. 
One intriguing aspect in this development 
is the saga of the $\Delta I=1/2$ rule: formulated in a compact way 
\cite{DELTARULE} it 
was originally expected to find a simple dynamical explanation; several 
enhancement factors were indeed found, but the observed enhancement 
could not be reproduced in a convincing manner; this problem was then 
bracketed for some future reconsideration and it was argued that 
$\epsilon ^{\prime}/\epsilon$ could be predicted while ignoring the 
$\Delta I=1/2$ rule. Some heretics -- `early' ones  
\cite{SANDA} and `just-in-time' ones \cite{TRIESTE} -- 
however argued that only approaches that reproduce the observed 
$\Delta I=1/2$ enhancement can be trusted to yield a re alistic 
estimate of $\epsilon ^{\prime}/\epsilon$. In particular 
it had been suggested that the scalar $\pi\pi$ resonance 
called $\sigma$  
plays a significant role here \cite{SANDA}. 

In all fairness one 
should point out that due to the large number of contributions with 
different signs theorists are facing an unusually complex 
situation \cite{FABB}. One can hope for
lattice QCD  to come through, yet it has to go beyond the 
quenched approximation, which will require more time.  

The second new element in 1999 was the start-up of the 
new asymmetric $B$ factories BaBar and BELLE. 
Their first results again leave us 
in limbo \cite{GOLUT}:
\bea 
{\rm sin}2\phi _1[\beta] &=& 0.45 ^{+0.43 + 0.07}_{-0.44 - 0.09}  \; \; 
{\rm BELLE} \\
{\rm sin}2\phi _1[\beta] &=& 0.12 \pm 0.37 \pm 0.09  \; \; 
{\rm BaBar}
\eea 
to be compared with the earlier data
\beq 
{\rm sin}2\phi _1[\beta] = 0.79 \pm 0.44 \;    
{\rm CDF}
\eeq
It is natural to ask what we would learn 
from a `Michelson-Morley outcome', if, say, 
$|{\rm sin}2\phi_1| < 0.1$ were established. Firstly, we would know 
that the KM ansatz would be ruled out as a major player 
in $K_L \to \pi \pi$ -- there would be no plausible deniability! 
Secondly, one would have to raise the basic question why the 
CKM phase is so suppressed, unless there is a finely tuned cancellation 
between KM and New Physics forces in $B\to \psi K_S$; this would shift 
then the CP asymmetry in $B\to \pi \pi, \, \pi \rho$. 

I expect those 
$B$ factories to have established CP violation in at 
least one $B$ decay mode by 2002. Yet that will not be the end of it -- 
far from it! Experiments at the upgraded $B$ factories at KEK and 
SLAC together with new experiments at the LHC -- LHC-B -- and at 
FNAL -- BTeV -- are expected to achieve experimental accuracies of 
a few percent, and they will measure many more observables. 
At the same time I expect that over the next five years or so 
we will be able to predict Standard Model effects with a few percent 
accuracy due to the improved theoretical tools 
sketched above and new 
measurements of CP {\em in}sensitive rates. We will then face 
the following type of challenge: how confident will we be in 
inferring the intervention of New Physics based on a difference 
between data and predictions? 

In principle there are precedents for establishing the presence of 
New Physics in such an {\em indirect} way in heavy flavour 
decays: based on the apparent absence of flavour changing neutral 
currents some courageous souls 
\cite{GIM} postulated the existence of charm quarks; 
the occurance of $K_L \to \pi \pi$ lead to the conjecture that 
even a third family of quarks had to exist \cite{KM}. However in 
all those cases we could rely on a {\em qualitative} 
discrepancy; i.e., the difference between observed and predicted rate 
amounted to several orders of magnitude or the predicted 
rate was zero -- as for $K_L \to \pi \pi$. In the decays of beauty 
hadrons we predict many large or at least sizeable effects, and 
realistically in most cases we can expect differences well below an 
order of magnitude only! 
E.g., one predicts an asymmetry of, say, 40 \%, but observe 
-40\%: will we all be confident enough to claim the presence of 
New Physics then? What about 40\% vs. 60 \% or even vs. 50\%? This 
would represent a novel challenge not encountered before; it will 
require that we gain quantitative control over that most evasive 
class of entities -- theoretical uncertainties. I am confident 
we will make great progress in that respect. My optimism is 
not based on hoping that novel 
theoretical breakthroughs will occur although they might. 
But what will empower us is the 
fact that so many different types of observables can be measured 
in beauty decays. There are actually six KM unitarity triangles 
\cite{TRIANGLES}, and several of their angles can be 
measured in the dedicated and comprehensive research program 
that is being undertaken world-wide. Our analysis will then 
be able to invoke overconstraints -- the most effective weapon 
in our arsenal against systematic uncertainties in general!

%%%%%%%%%%%%%%%%%%%%%%%%
\section{Status of T and CPT Invariance}
%%%%%%%%%%%%%%%%%%%%%%%%

It is often 
alleged that 
CPT invariance can boast of impressive experimental verification 
as expressed 
through the bound 
$|M(K^0) - M(\bar K^0)|/M(K) = (0.08 \pm 5.3) \cdot 10^{-19}$. 
However one might as well have divided this difference by the mass of an 
elephant since {\em intrinsically} the kaon mass is only 
marginally more 
related to the $K - \bar K$ mass splitting than the elephant's mass.

To put it differently: since this CPT breaking is expressed through 
a mass difference, one needs another 
{\em dimensional} quantity as yardstick.  This can be provided by 
Im$M_{12}$ expressing CP violation in the mass matrix: 
\beq 
|M(K^0) - M(\bar K^0)| < 2.5 \cdot 10^{-10} \; {\rm eV} 
\; \; \Leftrightarrow \; \; 
{\rm Im}M_{12} \simeq 10^{-8} \; {\rm eV} \; ; 
\eeq
i.e., CPT breaking still could be as `large' as a few percent of the 
observed CP violation! 

I have similar reservations about expressing bounds on the 
mass difference between protons and antiprotons relative to the 
proton mass etc.

Similarly I find statements relating bounds on the 
mass difference between protons and antiprotons to the proton mass 
as merely mathematical and largely devoid of physical meaning. 

I want to emphasize that our belief in 
CPT invariance is based much more on `dogma', i.e. theory, 
than empirical facts. For it 
is an almost inescapable consequence of {\em local} 
quantum field theories based on canonized assumptions like 
Lorentz invariance, the existence of a unique vacuum state and 
weak local commutativity obeying the `right' statistics. 
Some explicit examples of CPT breaking theories have been 
given, but they are highly contrived and unattractive 
\cite{TOD,TUMB}. 

The new interest in experimental studies of CPT symmetry is 
fed by two more recent developments 
\cite{KOST}: 
\begin{itemize}
\item 
Novel tests of CPT as well as {\em linear} quantum mechanics 
can be performed at the $\Phi$ and beauty factories DA$\Phi$NE, 
BABAR and BELLE respectively by harnessing EPR correlations 
\cite{EPR}. 
\item 
Superstring theories are intrinsically {\em non}local thus vitiating 
one of the central axioms of the CPT theorem. Furthermore gravity 
could induce CPT breaking either as a true symmetry violation or as 
a background effect due to the preponderance of matter over antimatter 
in our corner of the universe. Then it would be not unreasonable to 
expect CPT asymmteries to scale like a positive power of 
$E/M_{Planck}$. {\em If} that power were unity one would guestimate 
$|M(K^0) - M(\bar K^0)| \sim M(K)/M_{Planck} \sim 10^{-19}$; yet 
the main argument in favour of such a scenario is `why not?".

\end{itemize}
Unfortunately these suggestions do not yield any reliable benchmark 
figures for CPT violations. Searches for them still represent 
shots in the dark, although there is a wide field for them  
\cite{RUSSELL}.   

In this context there might be more interest in the even more 
unorthodox  
suggestion that the extra dimensions 
required by superstring theories are larger than 
the Planck length by many orders of magnitude  
\cite{HALL}. This leads to the intriguing scenario where the 
Planck scale is actually a derived rather than a fundamental one;  
that role is played by a much lower energy scale $M_X$. 
It was noticed that the $1/r^2$ force law 
for gravity had not been tested in the sub-millimeter domain. 
With gravity (unlike gauge forces) 
operating in {\em all} dimensions, their dynamics 
would undergo a great qualitative change at distances 
comparable to the size of the extra dimensions. 
More specifically the $1/r^2$ law would change to 
$1/r^{2 + n}$ with the natural number $n$ depending on 
the number of extra dimensions and the new fundamental 
unification scale $M_X$. In such a case it just might be conceivable 
that studying the spectra of {\em anti-protonic} atoms 
could reveal an {\em apparent}  
violation of CPT symmetry. The picture I have in mind without having 
done a calculation that is certainly doable is the following: in 
an anti-protonic atom where the antiproton   
is as close to the nucleus as possible without 
entering the meson cloud around the latter the orbiting 
antiproton would experience a gravitational force exceeding 
the canonical one by many orders of magnitude if $n$ were 
sufficiently large. Its gravitational mass could then 
differ significantly from the mass of protons {\em determined 
at larger distances}.  

Although CP violation implies T violation due to the CPT theorem 
(and despite my skepticism concerning the observability 
of the latter), 
I consider it
highly significant that more  direct evidence has been obtained through the 
`Kabir test': CPLEAR has found \cite{CPLEAR} 
\beq 
A_T \equiv 
\frac{\Gamma (K^0 \to \bar K^0) - \Gamma (\bar K^0 \to K^0)}
{\Gamma (K^0 \to \bar K^0) + \Gamma (\bar K^0 \to K^0)} = 
(6.6 \pm 1.3 \pm 1.0)\cdot 10^{-3} 
\eeq
versus the value $(6.54 \pm 0.24)\cdot 10^{-3}$ inferred from 
$K_L \to \pi^+\pi^-$. Of course, some assumptions still 
have to be made, namely that {\em semileptonic} $K$ decays obey 
CPT or that the Bell-Steinberger relation is satisfied with 
{\em known} decay channels only. Avoiding both assumptions 
one can write down an 
admittedly contrived scheme where the CPLEAR data are  
reproduced {\em without} T violation; the price one pays is a large CPT 
asymmetry $\sim {\cal O}(10^{-3})$ in 
$K^{\pm} \to \pi ^{\pm}\pi ^0$ \cite{TBS}.

KTeV and NA48 have analyzed the rare decay 
$K_L \to \pi^+\pi^- e^+e^-$ and found a large {\em T-odd} 
correlation between the $\pi^+\pi^-$ and $e^+e^-$ planes in 
full agreement with predictions \cite{SEHGAL}. 
Let me add just two comments here: (i) This agreement cannot be 
seen as a success for the KM ansatz. Any scheme reproducing  
$\eta _{+-}$ will do the same. (ii) The argument that strong final state 
interactions (which are needed to generate a T odd correlation 
above 1\% with T invariant dynamics) cannot affect the 
relative orientation 
of the $e^+e^-$ and $\pi ^+\pi ^-$ planes fails on the 
quantum level \cite{TBS}. 
 
The effect found represents a true CP asymmetry. Yet if one is 
sufficiently determined, it still could be attributed to CP and 
CPT breaking that leaves T invariant. A more detailed discussion 
of these subtle points is given in \cite{SEHGAL,TBS}.

%%%%%%%%%%%%%%%%%%%%%%
\section{Beyond the Mainstream}
%%%%%%%%%%%%%%%%%%%%

Considerable circumstantial evidence has been accumulated 
that the SM is incomplete. There are (at least) four central mysteries 
at the basis of flavour dynamics: 
\begin{itemize}
\item 
Why is there a family 
structure relating quarks and leptons? 
\item 
Why is there more than 
one family, why three, is three a fundamental parameter? 
\item 
What is the origin of the observed pattern in the quark masses 
and the  
CKM parameters? This pattern can hardly have come about by accident. 
\item 
Why are neutrinos massless -- or aren't they? 
\end{itemize}
To a large degree studying flavour dynamics represents an indirect 
or high sensitivity search for New Physics, as already stressed 
in my discussion of $B$ physics. Yet we have to be 
sufficiently openminded in where we look; i.e., search also in 
areas where the Standard Model does not predict observable 
effects. 

One such area is represented by searching for direct CP asymmetries 
in hyperon decays where the SM effects are below the 
sensitivity level of the ongoing HyperCP experiment. 

Others are even more radical and can be characterized as a 
`King Kong' scenario: "One might be unlikely to encounter King 
Kong; yet once it happens there can be no doubt that one has 
come across someting out of the ordinary". Such a situation 
can be realized for $K_{\mu 3}$ decays and EDMs -- as introduced 
in Sect. 2.2 -- and to 
some degree for charm transitions. 

%%%%%%%%%%%%
\subsection{$P_{\perp}(\mu )$ in $K^+ \to \mu ^+ \pi ^0 \nu$}
%%%%%%%%%%%%%%

With $P_{\perp}(\mu ) \sim 10^{-6}$ in the SM, it would also 
reveal New Physics that has to involve chirality breaking weak 
couplings: $P_{\perp}(\mu ) \propto {\rm Im}\xi$, where 
$\xi \equiv f_-/f_+$ with 
$f_-[f_+]$ denoting the chirality violating [conserving] 
decay amplitude. There is an on-going experiment at KEK 
(KEK-E 246) aiming at 
a sensitivity for $P_{\perp}(\mu )$ of $10^{-3}$ or better.

%%%%%%%%%%
\subsection{EDM's}
%%%%%%%%%%%

With the KM scheme predicting unobservably tiny effects 
(with the only exception being the `strong CP' problem) --  
namely $d_{N,e} < 10^{-30}$ e cm --  
and many New Physics scenarios yielding 
$d_N$, $d_e$ $\geq 10^{-27}$ ecm, this is truly a 
promising zero background search for New Physics! The next round of 
experiments is aiming at $10^{-28}$ e cm for $d_N$ and 
$10^{-30}$ e cm for $d_e$ \cite{HINDS}. 

The game one is hunting is actually much more numerous, since many 
effects from the domain of nuclear physics can be employed here
\cite{HINDS}.

%%%%%%%%%%%%%
\subsection{$D^0$ Oscillations \& CP Violation}
%%%%%%%%%%%%%%%

It is often stated that $D^0$ oscillations are slow and 
CP asymmetries tiny within the SM and that therefore their analysis 
provides us with zero-background searches for New Physics. 

Oscillations are described by the normalized mass and width 
differences:  
$x_D \equiv \frac{\Delta M_D}{\Gamma _D}$,   
$y_D \equiv \frac{\Delta \Gamma}{2\Gamma _D}$.  
A conservative SM estimate yields $x_D$, $y_D$ $\sim 
{\cal O}(0.01)$. Stronger bounds have appeared in the literature, 
namely that the contributions from the 
operator product expansion (OPE) are completely insignificant 
and that long distance contributions {\em beyond} the OPE provide the 
dominant effects yielding $x_D^{SM}$, $y_D^{SM}$ 
$\sim {\cal O}(10^{-4} - 10^{-3})$. A recent detailed analysis 
\cite{BUOSC} revealed 
that a proper OPE treatment reproduces also such long distance 
contributions with  
\beq 
x_D^{SM}|_{OPE}, \, y_D^{SM}|_{OPE} \sim {\cal O}(10^{-3}) 
\eeq   
and that $\Delta \Gamma $, which is generated from  
on-shell contributions, is -- in contrast to $\Delta m_D$
-- insensitive to New Physics while on the other hand more susceptible 
to violations of (quark-hadron) duality. 

Four experiments have reported new data on $y_D$: 
\bea
 y_D &=& 
\left\{ 
\begin{array}{l} 
(0.8 \pm 2.9 \pm 1.0) \% \; \; {\rm E791} \\
(3.42 \pm 1.39 \pm 0.74) \% \; \; {\rm FOCUS} \\ 
(1.0^{+3.8 + 1.1}_{-3.5-2.1})\% \; \; {\rm BELLE}
\end{array}  
\right. 
\\  
y_D^{\prime} &=& (-2.5 ^{+1.4}_{-1.6} \pm 0.3)\% \; \; {\rm CLEO}
\eea
E 791 and FOCUS compare the lifetimes for two different channels, 
whereas CLEO fits a general lifetime evolution to 
$D^0(t) \to K^+\pi ^-$; its $y_D^{\prime}$ depends on the strong 
rescattering phase between $D^0 \to K^-\pi^+$ and 
$D^0 \to K^+\pi^-$ and therefore could differ substantially from 
$y_D$ if that phase were 
sufficiently large. 
The FOCUS data contain a suggestion that the lifetime 
difference in the 
$D^0 - \bar D^0$ complex might be as large as ${\cal O}(1\% )$. 
{\em If} $y_D$ indeed were $\sim 0.01$, two scenarios could arise 
for the mass difference. If $x_D \leq {\rm few} \times 10^{-3}$ 
were found, one would infer that the $1/m_c$ expansion yields a 
correct semiquantitative result while blaming the large value for 
$y_D$ on a sizeable and not totally surprising violation of 
duality. If on the other hand $x_D \sim 0.01$ would emerge, we would face 
a theoretical conundrum: an interpretation ascribing this to New 
Physics would hardly be convincing since $x_D \sim y_D$. A more sober 
interpretation would be to blame it on duality violation or on the 
$1/m_c$ expansion being numerically unreliable. Observing 
$D^0$ oscillations then would not constitute a `King Kong' 
scenario. 

Searching for {\em direct} CP violation in 
Cabibbo suppressed $D$ decays as a sign for New Physics would also 
represent a very complex challenge: within the KM description one expects 
to find some asymmetries of order 0.1 \%; yet it would be hard 
to conclusively rule out some more or less accidental enhancement due to a 
resonance etc. raising an asymmetry to the 1\% level. 

The only clean environment is provided by CP violation involving 
$D^0$ oscillations, like in $D^0(t) \to K^+ K^-$ and/or 
$D^0(t) \to K^+ \pi ^-$. For the asymmetry would depend 
on the product sin$(\Delta m_D t) \cdot {\rm Im}
[T(\bar D\to f)/T(D\to \bar f)]$: with both factors being 
$\sim
{\cal O}(10^{-3})$ in the SM one predicts a practically zero 
effect. Yet New Physics scenarios can induce signals as large 
as order 1 percent for $D^0(t) \to K^+ K^-$ and even larger for 
$D^0(t) \to K^+ \pi^-$.

%%%%%%%%%%%%
\section{Outlook}
%%%%%%%%%%%%
I want to start with a statement about the past: 
{\em The comprehensive study of kaon and hyperon physics 
has been instrumental in guiding us to the Standard Model.}  
\begin{itemize}
\item 
The $\tau -\theta $ puzzle led to the realization that parity is not 
conserved in nature. 
\item 
The observation that the production rate exceeded the decay rate 
by many orders of magnitude -- this was the origin of the 
name `strange particles' -- was explained through postulating 
a new quantum number -- `strangeness' -- conserved by the strong, 
though not the weak forces. This was the beginning of the second 
quark family. 
\item 
The absence of flavour-changing neutral currents was incorporated 
through the introduction of the quantum number `charm', which 
completed the second quark family. 
\item 
CP violation finally led to postulating yet another, the third 
family. 
\end{itemize}
All of these elements which are now essential pillars of the Standard 
Model were New Physics at {\em that} time! 

I take this historical 
precedent as clue that a detailed, comprehensive and thus 
neccessarily long-term program on the dynamics of heavy flavours 
-- on the quark as well as lepton side -- in general and on CP 
violation in particular will lead to a 
new paradigm, a {\em new} Standard Model. For we are addressing 
the problem of fermion mass generation -- a 
central mystery in our present SM. Such studies are of fundamental 
importance, they will teach us lessons that cannot be obtained 
any other way and cannot become obsolete. 

It will not be an easy journey on a straight path, nor will it be short, 
nor can we anticipate where we will end up. Yet we know 
that we are at 
the beginning of an truly exciting adventure. 

Finally we should never 
loose sight of the fact that by any historical standard we are 
generously supported by the public. Therefore we better appreciate 
how highly privileged we are in participating in this adventure. 
I do not know of a sufficient justification for this privilege, 
only of a necessary one: to work with as much dedication as we can 
possibly muster and never be satisfied with a second-best effort!

\vskip 3mm  
{\bf Acknowledgements} 

While I have enjoyed meetings at many beautiful places in the world, 
I have to say that San Servolo is second to none. Its setting in the 
laguna of Venice, the views and sunsets it offers, how it combines its 
own tranquility with the nearby vibrancy of Venice, its sparkling light -- 
all of this is beyond description, it has to be experienced. Prof. 
Faessler deserve our enthusiastic thanks for introducing this jewel into the 
conference scene of particle physics. 
This work has been supported by the NSF under the grant 
PHY 96-05080. 

%%%%%%%%%%%%%%%%%%%%%%%%%%%%%%%

\end{document}